\documentclass
[aps,twocolumn,prl,showpacs,preprintnumbers,amsmath,amssymb]{revtex4}
\usepackage{natbib}
\usepackage[dvips]{graphicx}
\usepackage{dcolumn}
\usepackage{bm}
\usepackage{times}
\usepackage{mathptmx}

\begin{document}
\draft

\title{Randomly Diluted $e_g$ Orbital-Ordered Systems}
\author{T. Tanaka, M. Matsumoto, and S. Ishihara} 
\address{Department of Physics, Tohoku University, Sendai 980-8578, Japan}
\date{\today}

\begin{abstract}
Dilution effects on the long-range ordered state of the doubly degenerate $e_g$ orbital  
are investigated. 
Quenched impurities without the orbital degree of freedom are introduced 
in the orbital model where the long-range order is realized by the order-from-disorder mechanism. 
It is shown by the Monte-Carlo simulation and the cluster-expansion method that 
a decrease in the orbital-ordering temperature by dilution is remarkable 
in comparison with that in the randomly diluted spin models.
Tilting of orbital pseudo-spins around impurity is the essence of this dilution effects. 
The present theory provides a new view point for the recent resonant x-ray scattering 
experiments in KCu$_{1-x}$Zn$_x$F$_3$.
\end{abstract}


\maketitle
\narrowtext
Impurity effects on the long-range ordered state are one of the attractive themes in 
recent study of correlated electron systems \cite{book}. 
A small amount of non-magnetic impurities dramatically destroy 
the superconductivity in cuprates, and 
induce the antiferromagnetic long-range order in the low-dimensional quantum spin-liquids. 
Doped impurities also cause striking effects on the charge- and orbital-orders; 
substitution of Cr ions for Mn in colossal magnetoresistive (CMR) manganites immediately 
destroys the charge-orbital orders \cite{barnabe}. 
Origin of CMR itself is studied from the view point of randomness and/or percolation \cite{uehara}. 

Recently, the dilution effects in KCuF$_3$ by 
substituting Zn for Cu is reported experimentally by 
Tatami ${\it et\ al.}$~\cite{tatami}. 
A Cu$^{2+}$ ion in the cubic-crystalline field 
shows the $t_{2g}^6 e_g^3$ configuration 
which has the $e_g$ orbital degree of freedom. 
The Cu ions in the perovskite crystal  
form the three-dimensional (3D) simple-cubic (SC) lattice, 
and exhibit the long-range orbital order (OO) in room temperatures, 
where the $d_{y^2-z^2}$- and $d_{z^2-x^2}$-like orbitals 
are aligned with momentum ${\bf Q}=(\pi, \pi, \pi)$. 
Since the five $d$ orbitals are fully occupied in Zn$^{2+}$, 
substitution of Zn for Cu corresponds to dilution of orbital. 
The resonant x-ray scattering (RXS) studies in KCu$_{1-x}$Zn$_x$F$_3$ reveal that 
a decrease in the orbital-ordering temperature ($T_{\rm OO}$) by dilution is 
remarkable in comparison with the randomly diluted magnets, 
and OO disappears around $x=0.5$, as shown in the inset of Fig.~1. 
These observations may not be explained by the conventional percolation 
scenario; the site-percolation threshold in 3D SC 
lattice is $x_c(\equiv 1-p_c)=$0.69 which is applicable well to the several 
diluted magnets such as KMn$_{1-x}$Mg$_x$F$_3$\cite{sinchcombe,breed}.

\begin{figure}[tb]
\begin{center}
\includegraphics[width=\columnwidth,clip]{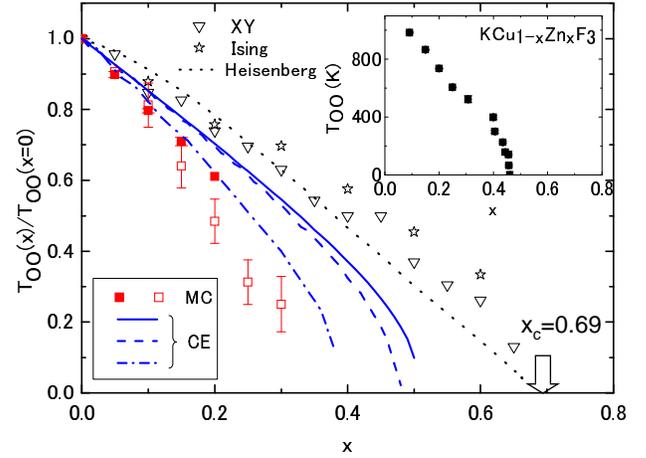}
\caption{Impurity concentration dependence 
of the normalized orbital-ordering temperature $T_{\rm OO}(x)/T_{\rm OO}(x=0)$. 
The calculated results by the MC method are 
shown by the closed- and open-red squares (see text), 
and those by the CE method for the one- and two-site clusters 
are shown by the bold- and broken-blue lines, respectively. 
The results by the CE method 
where the PS operator is treated as a classical vector are shown 
by the one-point chain line (blue). 
For comparison, 
$T_{\rm N}$ in the XY (reverse-triangles) and Ising (stars) models 
by the MC method, and 
that in the Heisenberg model (dotted line) by the CE method 
are also shown. 
The inset shows the Zn concentration dependence of $T_{\rm OO}$
in KCu$_{1-x}$Zn$_x$F$_3$ obtained by the RXS experiments \protect\cite{tatami}. 
}
\label{fig1}
\end{center}
\end{figure}

We examine, in this Letter, the dilution effects on the long-range order of 
the $e_g$ orbital degree of freedom. 
As well known, 
the doubly degenerate $e_g$ orbital is treated by the pseudo-spin (PS) operator with a magnitude of 1/2; 
${\bf T}_{i}=\frac{1}{2} \sum_{\gamma, \gamma', s} 
d_{i \gamma s}^\dagger 
{\bf \sigma}_{\gamma \gamma'}
d_{i \gamma' s}  
$
where $d_{i \gamma s}$ indicates the annihilation operator of a hole with spin 
$s$ and orbital $\gamma$ at site $i$,  
and ${\bf \sigma}$ is the Pauli matrices. 
A shape of the electronic orbital is described by an angle $\theta$ of the PS vector as 
$|\theta \rangle=\cos(\theta/2)|d_{3z^2-r^2} \rangle+\sin(\theta/2)|d_{x^2-y^2} \rangle$. 
For example, $\theta=0$, $2\pi/3$ and $4\pi/3$ correspond to
the $d_{3z^2-r^2}$, $d_{3x^2-r^2}$, and  
$d_{3y^2-r^2}$ orbitals, respectively,  
and $\theta=\pi/3$, $\pi$ and $-\pi/3$ to $d_{y^2-z^2}$, $d_{x^2-y^2}$, and $d_{z^2-x^2}$, respectively. 
As shown below, 
the Hamiltonian for the diluted orbital system is described by the PS operators 
and seems to be on the same footing with the spin models. 
However, the obtained results are qualitatively different from the diluted magnets. 
The main results are shown in Fig.~\ref{fig1}. 
The impurity-concentration $x$ dependences of $T_{\rm OO}$, calculated by the two different methods, 
show rapid decreases in comparison with that in the spin models. 
The results can explain the experimentally discovered anomalous dilution effects in 
KCu$_{1-x}$Zn$_x$F$_3$~\cite{tatami}. 

\par
The model Hamiltonian adopted here 
describes the orbital interaction between 
the nearest-neighbor (NN) Cu sites in a 3D SC lattice: 
\begin{equation}
{\cal H}=2J \sum_{\langle i j \rangle} \tau_i^l \tau_j^l  \varepsilon_i \varepsilon_j, 
\label{eq:tt}
\end{equation} 
where $J(>0)$ is the interaction, and 
$\langle ij \rangle$ indicates a pair of the NN sites 
along the direction $l(=x, y, z)$. 
The operator $\tau^l_i$ depends explicitly on the bond direction 
and is defined by a linear combination of the PS operators as 
\begin{equation}
\tau_i^l=\cos \left ( \frac{2\pi}{3} n_l \right ) T_{iz}
+\sin \left ( \frac{2\pi}{3} n_l \right ) T_{ix}  , 
\end{equation}
with $(n_x, n_y, n_z)=(1, 2, 3)$.
The quenched impurities without the orbital degree of freedom are introduced 
by $\varepsilon_i$ taking one and zero for Cu and Zn, respectively. 

The Hamiltonian in Eq.~(\ref{eq:tt}) without impurities ($\varepsilon_i=1$ for $\forall i$) 
does not concern an origin of the interaction, i.e. 
the electronic and/or phononic interactions. 
In the electronic view point, this is derived by the generalized-Hubbard model   
with the $e_g$ orbital degree of freedom. 
Through the perturbational expansion with respect to the NN electron transfer $t$, 
the spin-orbital superexchange model is obtained \cite{kugel,ishihara0}; 
$
{\cal H}_{ST}=-2J_1\sum_{\langle ij \rangle} 
( \frac{3}{4}+{\bf S}_i \cdot {\bf S}_j  )
( \frac{1}{4} - \tau^l_i \tau_j^l )
-2J_2 \sum_{\langle ij \rangle} 
( \frac{1}{4} - {\bf S}_i \cdot {\bf S}_j  )
( \frac{3}{4} + \tau^l_i \tau_j^l +\tau_i^l + \tau_j^l ) 
$. 
Here, ${\bf S}_i$ is the spin operator at site $i$ 
with magnitude of $1/2$, and 
$J_1(=t^2/(U-3I))$ and $J_2(=t^2/U)$ are the superexchange interactions with 
the on-site intra-orbital Coulomb interaction $U$ and 
the exchange interaction $I$. 
Since the N$\rm \acute e$el temperature ($T_{\rm N}$) in KCu$_{1-x}$Zn$_x$F$_3$ 
is far below $T_{\rm OO}$ in a whole range of $x$~\cite{tatami},  
taking ${\bf S}_i \cdot {\bf S}_j=0$ is a good assumption, 
and Eq.~(\ref{eq:tt}) with $J=\frac{3}{4}J_1-\frac{1}{4}J_2$ is obtained. 
In the phononic view point, 
the orbital model is derived based on the cooperative Jahn-Teller (JT) effects. 
Start from the linear JT coupling Hamiltonian 
$
{\cal H}_{JT}=g\sum_{i, m=(x, z)} Q_{i m} T_{i m}
$ 
with the vibrational modes $Q_{i m}$ $(m=x, z)$ in a F$_6$ octahedron,  
and the lattice potential $K$ between NN Cu-F bonds.  
After rewriting $Q_{im}$ by the phonon coordinates ${\bf q}_{\bf k}$, 
we obtain Eq.~(\ref{eq:tt}) with $J=2g^2/(9K)$ by 
integrating out ${\bf q}_{\bf k} $\cite{kanamori,okamoto}.  
A sign of $J$ is positive in both the two processes. 
A unique aspect of the orbital model to be noticed here 
is that the explicit form of the interaction depends on the bond direction $l$; 
the interactions are $T_{iz} T_{jz}$ for $l=z$, 
and $[-\frac{1}{2}T_{iz}+(-)\frac{\sqrt{3}}{2}T_{ix}]
[-\frac{1}{2}T_{jz}+(-)\frac{\sqrt{3}}{2}T_{jx}]$ for $l=x (y)$. 
When we focus on one direction $l$, 
the staggered alignment of the $d_{3l^2-r^2}$ and $d_{m^2-n^2}$ orbitals  
with $(l, m, n)=(x,y,z), (y,z,x)$ and $(z,x,y)$ is favored. 
In a 3D SC lattice with and without impurities,   
the stable orbital configurations are non-trivial. 

We have attacked this issue by the numerical approach together with the analytical one, i.e. 
the classical Monte-Carlo (MC) simulation, 
and the cluster expansion (CE) method. 
In the MC method, the PS operator is treated as a classical vector 
defined in the $T_x-T_z$ plane. 
The MC calculations have been performed for the cubic 
$L \times L \times L$ lattice ($L=10 \sim 18$) with the periodic-boundary condition. 
For each sample, 30,000MC steps are spent for measurement after 
8,000MC steps for thermalization. 
The physical quantities are averaged over $20 \sim 80$ MC samples at each parameter set. 
We adopt the CE method proposed in Ref.~\cite{mano}, 
where the fluctuations of the effective fields are determined with the order parameter,  
self-consistently. 
Even in the two-site clusters, 
this CE method provides good values for 
$x_c$ and the Curie temperature $T_{\rm C}$ at $x=0$ for Ising and Heisenberg models \cite{mano}. 

\begin{figure}[tb]
\begin{center}
\includegraphics[width=\columnwidth,clip]{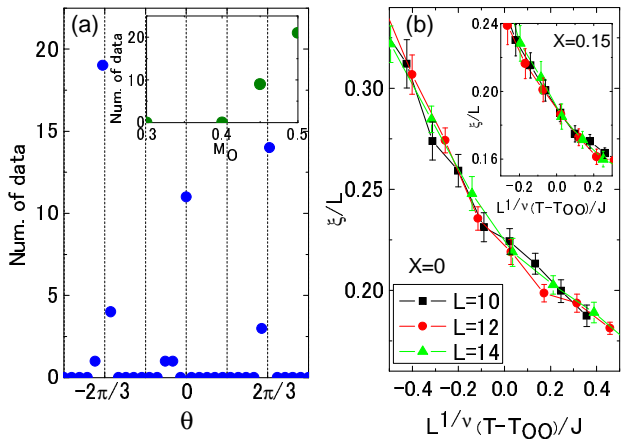}
\caption{
(a) A MC data distribution for the orbital angle $\theta$ in the 
staggered-type orbital order at $x=0$. 
The inset shows a data distribution of the order 
parameter $M_{\rm O}$ at $x=0$. 
(b) The finite-size scaling for the correlation length $\xi/L$ 
as functions of $L^{1/\nu}(T-T_{\rm OO})/J$ at $x=0$. 
The inset shows the scaling plot at $x=0.15$.}
\label{fig2}
\end{center}
\end{figure}

First, we show the results without impurities. 
It is known that, in the orbital model [Eq.~(\ref{eq:tt})] at $x=0$, 
there is a macroscopic number of degeneracy in the mean-field (MF) 
ground state \cite{feiner,khaliullin,ishihara2,kubo,nussinov}. 
These are classified into the two; 
(i)  
One of the MF solutions is the staggered-type OO with two 
sublattices, termed $A$ and $B$, and the momentum ${\bold Q}=(\pi, \pi, \pi)$. 
The orbital angles in the sublattices are $(\theta_A/\theta_B)=(\theta/\theta+\pi)$ 
for any value of $\theta$. 
Such continuous-rotational symmetry is unexpected 
from Eq.~(\ref{eq:tt}). 
(ii) 
Consider an OO with 
${\bf Q}=(\pi, \pi, \pi)$ and $(\theta_A/\theta_B)=(\theta_0/\theta_0+\pi)$, 
and focus on one direction in a 3D SC lattice, e.g. the $z$ direction. 
The MF energy is not changed by changing 
$ (\theta_0/\theta_0+\pi) \rightarrow (-\theta_0/-\theta_0-\pi)$ in each layer in the $xy$ plane.  
%
Both the types of degeneracy are understood from 
the momentum representation of Eq.~(\ref{eq:tt}):  
${\cal H}=\sum_{\bf k} \psi^t_{\bf k} \hat J({\bf k}) \psi_{\bf k}$ 
with $\psi_{{\bf k}}=[T_{z {\bf k}} , T_{x {\bf k}}]$ and the $2 \times 2$ matrix $\hat J({\bf k})$. 
By diagonalizing $\hat J({\bf k})$, we obtain the eigen values 
$ J_{\pm}({\bf k})=2J 
[-c_x-c_y-c_z \pm \sqrt{c_x^2+c_y^2+c_z^2-c_x c_y-c_y c_z-c_z c_x}  ] $ 
with $c_l=\cos (\pi k_l/a)$ \cite{ishihara3}. 
$J_-({\bf k})$ has its minimum along $(\pi, \pi, \pi)-(0, \pi, \pi)$ and 
other three-equivalent directions. 
A lifting of the degeneracy in the orbital model 
is discussed in Refs.~\cite{kubo, nussinov} by the spin-wave analyses. 
Here we show the lifting of the degeneracy by the MC method. 
In Fig.~\ref{fig2} (a), number of data obtained in the simulation 
is plotted as a function of the orbital angle $\theta$ in the 
staggered-type OO with $(\theta_A/\theta_B)=(\theta/\theta+\pi)$. 
It is shown that the type-(i) orbital degeneracy  
is lifted; $\theta$ is confined to the three discontinuous values of 
$0$, $2\pi/3$ and $4\pi/3$.   
The inset shows number of data as function of 
the orbital-order parameter $M_{\rm O}$ at ${\bf Q}=(\pi, \pi, \pi)$
defined as 
$M_{\rm O}^2=\sum_{m=x, z} \langle T_{{\bf Q} m} T_{-{\bf Q} m} \rangle$ 
with 
$T_{{\bf k} m}=\frac{1}{N}\sum_{i} e^{i {\bf k} \cdot {\bf r}_i} T_{im}$.  
Most of the data show $M_{\rm O}=0.5$, implying that 
the ordering pattern is of a staggered type, and 
the type-(ii) degeneracy is also lifted. 

We have performed  
the finite-size scaling analyses in the MC simulation to determine $T_{\rm OO}$. 
The correlation length $\xi$ 
is calculated by the second-moment method 
in the PS correlation function for several sizes $L$. 
In Fig.~\ref{fig2}(b), we demonstrate the scaling plot for $\xi/L$ as a function of  
$(T-T_{\rm OO})L^{1/\nu}$ at $x=0$. 
The scaling analyses work quite well for $L=$10, 12 and 14. 
$T_{\rm OO}$ and the exponent $\nu$ are determined 
through the least-square fitting by the polynomial expansion 
and obtained as  $T_{\rm OO}/J= 0.344 \pm 0.002$ and $\nu=0.69-0.81$, although 
statistical errors are 
not enough to obtain the precise value of $\nu$.   
Even in the diluted case [see the inset of Fig.~\ref{fig2}(b)], 
the precision is enough to determine $T_{\rm OO}[=(0.248 \pm 0.003)J$ for $x=0.15$ ]. 
Beyond $x=0.15$, the scaling analyses do not work well. 

\begin{figure}[tb]
\begin{center}
\includegraphics[width=\columnwidth,clip]{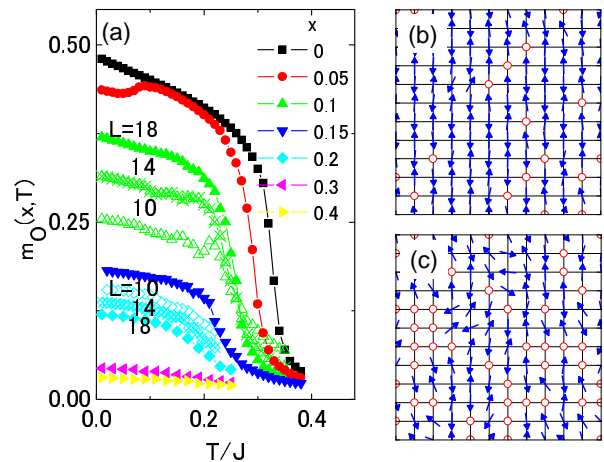}
\caption{
(a) Temperature dependence of the normalized order parameter 
$m_{\rm O}(x, T)[=M_{\rm O}/(1-x)]$ for various $x$ and $L$.  
(b) A snapshot in the MC simulation for the PS configuration at $x=0.1$. 
Open circles indicate the impurities. 
(c) A snapshot at $x=0.3$. }
\label{fig3}
\end{center}
\end{figure}

In Fig.~\ref{fig3}(a), the temperature dependence of the normalized-order parameter 
$\frac{1}{1-x}M_{\rm O}[\equiv m_{\rm O}(x,T)]$ are presented.
First, focus on the region of $x \leq 0.15$. 
As expected, 
$m_{\rm O}(x=0,T)$ abruptly increases at $T_{\rm OO}$ and 
is saturated to 0.5 at $T \rightarrow 0$. 
By doping impurity, 
$m_{\rm O}(x,T)$ does not reach 0.5 even far below $T_{\rm OO}$,  
and $m_{\rm O}(x \ne 0, T \rightarrow 0)$ gradually decreases with increasing $x$. 
Although the system sizes are not sufficient 
to estimate $m_{\rm O}(x,T)$ in the thermodynamic limit, 
$m_{\rm O}(x \ne 0,T \rightarrow 0)$ does not show the smooth 
convergence to 0.5 in contrast to the diluted spin models. 
Beyond $x=0.15$, the situation is changed qualitatively; 
in spite of the fact 
that $m_{\rm O}(x, T)$ grows around a certain temperature (e.g. $\sim 0.18J$ at $x=0.2$), 
$m_{\rm O}(x, T \rightarrow 0)$ becomes small abruptly and 
decreases weakly with increasing $L$. 
Anomalies in the specific heat 
and the susceptibility around this temperature become weak, and  
some of the physical quantities, e.g. $ N_{\rm O} \equiv \sum_{m={x,z}} \langle T_{ {\bf Q} m }\rangle^2 $ 
with ${\bf Q}=(\pi, \pi, \pi)$, 
depend on the initial 
states in the MC simulation. 
It seems that 
the orbital correlation grows up below this temperature, but 
the long-range order does not. 
We suppose that a possible orbital state in this region  
is a short-range ordered state or a glass state~\cite{binder}, 
although we do not identify this phase correctly at the present stage. 
Thus, in $x > 0.15$,  
a temperature where $dm_{\rm O}(x,T)/dT$ in $L=18$ takes a maximum is 
interpreted to be the cross-over and/or glass-transition temperature. 
Above $x=0.3$, estimation of $T_{\rm OO}$ becomes severe, 
because of the weak temperature dependence of $m_{\rm O}(x,T)$. 
Thus, as a supplementary information of the ordering temperature, 
we calculate a temperature $\widetilde T_{\rm OO}$ where $d N_{\rm O}/dT$ takes a maximum, 
and consider its full width at half maximum as an error. 
The initial state in the calculation for $\widetilde T_{\rm OO}$ is assumed to be the ordered state, 
although the initial-state dependence of $\widetilde T_{\rm OO}$ is much weaker than  
that of the $N_{\rm O}$ amplitude. 

The $x$ dependences of $T_{\rm OO}$ (and $\widetilde T_{\rm OO}$) 
presented in Fig.~\ref{fig1} are obtained by the 
MC (filled-red squares) and CE (blue lines) methods introduced above. 
For comparison, 
the $x$ dependences of $T_{\rm N}$ in the spin models are plotted. 
$\widetilde T_{\rm OO}$s are also shown by the open-red squares.
Starting to dope impurities in the orbital model, 
decrease of $T_{\rm OO}$ is more remarkable than that 
of $T_{\rm N}$ in the spin models. 
Around $x=0.15$, the clear transition 
to the long-range order becomes 
obscure, explained above. 
$\widetilde T_{\rm OO}$s obtained by $N_{\rm O}$ are close to $T_{\rm OO}$ below $x=0.2$,  
and decrease smoothly up to $x=0.3$, 
although the critical $x$, where $\widetilde T_{\rm OO}$ disappears, 
is not determined due to large statistical errors. 
A rapid decrease of $T_{\rm OO}$ is also obtained by the CE method. 
It is seen that difference between 
the calculated $T_{\rm OO}-x$ curves for the cluster sizes $N_{\rm C}=1$ (bold line) and 2 (broken like)
is within a few percent. 
$T_{\rm OO}$ in the CE method disappears around $x=0.5$ 
being much smaller than $x_c=0.69$ for the 3D SC lattice. 
We also calculate $T_{\rm OO}$ in the CE method where 
the quantum PS operators are replaced by the classical vectors. 
The calculated results plotted by the one-point chain line 
are close qualitatively to the $T_{\rm OO}({\widetilde T}_{\rm OO})-x$ curve obtained 
by the classical MC method. 

Now we introduce the physical picture of the diluted orbital-ordered state. 
The real-space configurations of the orbital PSs 
help us to understand the calculated $T_{\rm OO}-x$ curve.  
The MC snapshots of the PSs on a plane  
are shown in Figs.~\ref{fig3}(b) and (c) for $x$=0.1 and 0.3, respectively. 
The staggered-type OO with the orbital angles 
$(\theta_A/\theta_B)=(0/\pi)$ is seen in the back-ground of Fig.~\ref{fig3}(b). 
At the neighboring sites of the impurities indicated by the open circles, 
tiltings of the PS vectors are observed. 
Disturbing the PS configuration from $(0/\pi)$ becomes violent at $x=0.3$. 
These observations are caused by the local-symmetry breaking by dilution. 
Consider the orbital state at a neighboring site of an impurity on the $x$ axis. 
On account of the impurity, 
one of the interactions along $x$ vanishes. 
Since the interaction depends on the bond direction explicitly in the orbital model, 
the PS at this site tilts to gain the interaction energies for other five bonds. 
This is the essence of the diluted orbital systems and 
is in contrast to the conventional diluted spin models. 
To check roles of this PS tilting on the anomalous $T_{\rm OO}-x$ curve, 
we have performed the MC calculations in the model where 
the available orbital angles in the $T_z-T_x$ plane are restricted to a few states. 
The obtained $T_{\rm OO}-x$ curve is similar to that in the spin models; 
the tilting of PSs is confirmed to be the origin of the 
rapid decrease of $T_{\rm OO}$. 

\begin{figure}[tb]
\begin{center}
\includegraphics[width=\columnwidth,clip]{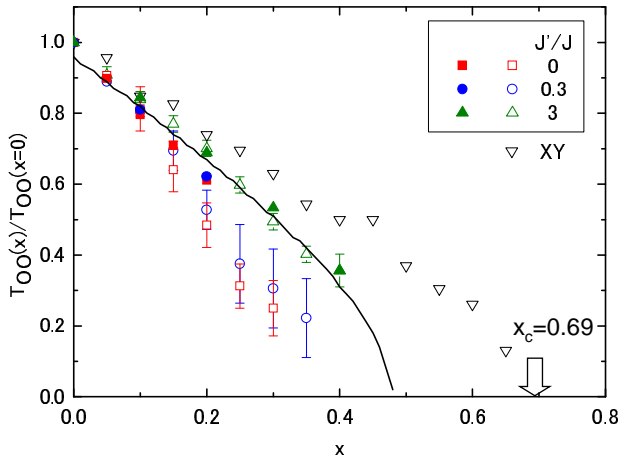}
\caption{
Impurity concentration dependence 
of the normalized orbital-ordering temperature $T_{\rm OO}(x)/T_{\rm OO}(x=0)$ 
with the higher-order JT coupling. 
The squares, circles and triangles indicate $T_{\rm OO}(x)/T_{\rm OO}(x=0)$ at $J'/J=$0, 0.3 and 3, respectively. 
$T_{\rm OO}$s indicated by the open and filled symbols are 
defined in the text. 
For comparisons, $T_{\rm N}$ of the XY model (reverse triangles), 
and $T_{\rm OO}$ at $J'/J$=0 by the CE method (bold line) are also plotted. }
\label{fig4} 
\end{center}
\end{figure}
Finally, to compare the present theory with the experiments in KCu$_{1-x}$Zn$_x$F$_3$, more directly, 
we introduce the higher-order JT coupling. 
From the RXS experiments~\cite{tatami} and the lattice 
distortion in KCuF$_3$~\cite{hirakawa}, 
the OO in KCuF$_3$ is expected to be the $d_{y^2-z^2}/d_{z^2-x^2}$-type; 
the cant-type OO with 
the orbital angles $(\theta/-\theta)$ ($\theta \sim \pi/3$)  
and ${\bf Q}=(\pi, \pi, \pi)$. 
To reproduce this type of OO, the higher-order JT coupling is required,  
${\cal H}_{HJT}=g'\sum_i \left \{
\left ( Q_{iz}^2-Q_{ix}^2 \right )T_{iz}-2Q_{iz}Q_{ix}T_{ix} 
\right \} 
$ 
which provides the anisotropy in the PS space. 
Here we treat approximately this term  
based on the theory of the cooperative JT effects. 
By rewriting $Q_{i x(z)}$ by the phonon coordinates ${\bf q}_{\bf k}$, 
and integrating out ${\bf q}_{\bf k}$, 
the interaction between orbitals at the different three-neighboring sites is obtained:   
$
{\cal H'}=J' \sum_{\langle ijk \rangle } \left \{ 
\left (T_{iz} T_{jz}+T_{ix}T_{jx} \right ) T_{kz}
-2 T_{iz}T_{jz}T_{kx}
\right \} 
\varepsilon_i \varepsilon_j \varepsilon_k 
$. 
Here, $J'=(16 g' g^2)/(9K^2) $ is the coupling constant 
and $\langle ijk \rangle$ indicates a sum of the neighboring three sites. 
The MF energy of ${\cal H'}$
is proportional to $\frac{J'}{8}  \cos 3 \theta$ 
implying the anisotropy.  
A value of $g'$ for a Cu$^{2+}$ ion 
was estimated by the adiabatic potential barrier in a molecule~\cite{shashkin} 
and corresponds to $J'$ being about $0.3J$. 
We have performed the MC calculations in the model of ${\cal H}+{\cal H}'$, 
and observed that the cant-type OO at $x=0$ is realized. 
The doping dependences of $T_{\rm OO}$ (and ${\widetilde T}_{\rm OO}$) with including ${\cal H'}$ 
are presented in Fig.~\ref{fig4} for $J'/J=0.3$ and 3.  
The filled and open symbols are for $T_{\rm OO}$ and $\widetilde T_{\rm OO}$, respectively, 
which are obtained by the same ways with those in Fig.~1. 
With increasing $J'$, the $T_{\rm OO}(\widetilde T_{\rm OO})-x$ curves approach to those for the spin models, 
because the anisotropy suppresses the tilting of PSs. 
However, we confirm that the rapid reduction of $T_{\rm OO}$ by dilution 
survives even in the calculation with the realistic value of $J'$, 
and is consistent with the RXS experimental results shown in the inset of Fig.~1. 

In summary, we have investigated the dilution effects on the long-range order of the $e_g$ orbital degree 
of freedom. 
We confirm, by both the MC and CE methods, that 
$T_{\rm OO}$ decreases rapidly with doping, in comparison 
with the diluted magnets. 
The tilting of the orbital PSs around impurities, which is distinguished qualitatively from the spin models,  
is the essence for the rapid decrease of $T_{\rm OO}$ .  
The present theory provides a new view point 
for the recent experiments in a diluted orbital systems of KCu$_{1-x}$Zn$_x$F$_3$. 
A broad peak profile observed by the RXS experiments~\cite{tatami} may be 
attributed to the orbital tilting around impurities. 
Observations of the tilting by 
the scanning-tunneling microscope and/or the scanning-electron microscope 
are a direct check for the present results. 
\par
Authors would like to thank Y.~Murakami for valuable discussion and 
permission to use the experimental data before publication. 
This work was supported by KAKENHI from MEXT, NAREGI and CREST. 
Part of the numerical calculation was performed by 
the supercomputers in IMR, Tohoku Univ., 
and ISSP, Univ. of Tokyo. 
T.~T. appreciates a financial support 
from Research Fellowship of JSPS. 
M.~M. thanks S.~Todo for helpful discussions. 
\vfill
\eject
\end{document}